\documentclass[prd,aps,showpacs,epsf,floats,onecolumn]{revtex4}%
\usepackage{amssymb}
\usepackage{amsfonts}
\usepackage{amsmath}
\usepackage{graphicx}%
\providecommand{\U}[1]{\protect\rule{.1in}{.1in}}
\begin{document}
\title{\textbf{Geometric aspects of analog quantum search evolutions}}
\author{\textbf{Carlo Cafaro}$^{1}$, \textbf{Shannon Ray}$^{2}$, and \textbf{Paul
M.\ Alsing}$^{2}$}
\affiliation{$^{1}$SUNY Polytechnic Institute, 12203 Albany, New York, USA}
\affiliation{$^{2}$Air Force Research Laboratory, Information Directorate, 13441 Rome, New
York, USA}

\begin{abstract}
We use geometric concepts originally proposed by Anandan and Aharonov to show
that the Farhi-Gutmann time optimal analog quantum search evolution between
two orthogonal quantum states is characterized by unit efficiency dynamical
trajectories traced on a projective Hilbert space. In particular, we prove
that these optimal dynamical trajectories are the shortest geodesic paths
joining the initial and the final states of the quantum evolution. In
addition, we verify they describe minimum uncertainty evolutions specified by
an uncertainty inequality that is tighter than the ordinary time-energy
uncertainty relation. We also study the effects of deviations from the time
optimality condition from our proposed Riemannian geometric perspective.
Furthermore, after pointing out some physically intuitive aspects offered by
our geometric approach to quantum searching, we mention some practically
relevant physical insights that could emerge from the application of our
geometric analysis to more realistic time-dependent quantum search evolutions.
Finally, we briefly discuss possible extensions of our work to the geometric
analysis of the efficiency of thermal trajectories of relevance in quantum
computing tasks.

\end{abstract}

\maketitle

\section{Introduction}

From a quantum mechanical perspective, one can modify a given state into
another state by acting upon the system with a convenient Hamiltonian. In
quantum computing, in particular, it is generally beneficial to know the path
connecting the two states in the shortest time with the maximum speed of
quantum evolution. Clearly, if the evolution occurs always at the maximum
speed, one achieve time optimality by transitioning from the initial to the
final state in the shortest time by taking the shortest route. Thus, the
problem of connecting these quantum states can be recast in the very
convenient form of a geodesic problem. From a more practical thermodynamical
perspective, high speed values are not beneficial when dealing with practical
devices that operate in finite time. Given the universal character of
thermodynamics laws, this latter fact remains valid also for actual
realizations of quantum computers. More specifically, high speeds lead to high
frictional losses which, in turn, hamper the thermal efficiency of these
physical systems similar to heat engines. Unfortunately, as a popular proverb
says, nothing comes from free: Time optimality and thermal efficiency occur in
conflicting favorable conditions. The former recommends high speeds in order
to shorten the duration of the physical process. The latter, instead, welcomes
low speeds to mitigate possible dissipative effects present in the system.
Achieving both time optimality and thermal efficiency is a very relevant and
challenging unresolved issue in quantum algorithm design \cite{castelvecchi97}%
. Given our awareness of the importance that geometric ideas play in physics,
acknowledging the power of regarding time optimality problems as geodesics
problems, and given this apparently unavoidable trade-off between speed and
thermal efficiency in the design of quantum algorithms, we are motivated here
to study in geometric terms the efficiency of quantum evolutions of relevance
in continuous-time quantum searching \cite{farhi98} with the hope of also
providing some helpful insights into the rather complicated speed-efficiency
trade-off quantification in quantum science. In what follows, we shall
introduce in more detail the problem that we discuss, its motivation, and its
relevance irrespective of our broader underlying interest represented by the
speed-efficiency trade-off in quantum algorithms design.

In the framework of quantum search algorithms \cite{grover,nielsen}, a
geodesic path with respect to the Fubini-Study metric in the projective
Hilbert space $%
\mathbb{C}
P^{N-1}$, with $N\overset{\text{def}}{=}2^{n}$ being the dimensionality of the
complex Hilbert space $\mathcal{H}_{2}^{n}$ of $n$-qubit quantum states,
emerges as a curve traced by the output quantum state specifying Grover's
original quantum search scheme \cite{alvarez00,wadati01,cafaro17}. In
exploring for efficient quantum circuits, Riemannian geometric techniques have
been exploited to reformulate the problem of finding optimal circuits into the
geometric problem of finding the shortest geodesic path between two points in
the curved geometry of the special unitary group $\mathrm{SU}\left(  N\right)
$ \cite{nielsen06,brandt10}. In the search for time-optimal quantum control
protocols, differential geometry techniques have been employed to recast the
quantum brachistochrone problem (for instance, see Ref. \cite{carlini06}) of
finding a control protocol capable of taking the minimum time to achieve a
desired task (for instance, the generation of a desired unitary gate) into a
problem of finding a shortest geodesic path on the special unitary group
$\mathrm{SU}\left(  N\right)  $ \cite{lupo15}. Interestingly, a transition
from a quantum state to an orthogonal one can be regarded as the elementary
step of a computational process \cite{levitin82,levitin09}. Moreover, from an
intuitive Riemannian geometric viewpoint, the optimal way of finding a
solution to an arbitrary computational problem appears to happen by
\textquotedblleft free falling\textquotedblright\ along the shortest geodesic
curve connecting the (chosen) initial and (desired) final states on the
appropriate curved manifold that characterizes the specific problem being
analyzed \cite{nielsen06}.

The work that we present in this article takes into account three key ideas:
i) The reformulation of a time-optimal problem into a geodesic problem
\cite{lupo15}; ii) The consideration that the most elementary step of a
computational process can be described in terms of a quantum mechanical
transition between two orthogonal states \cite{levitin82,levitin09}; iii) The
intuition that optimal solutions of computational tasks can be geometrically
described in terms of shortest geodesic paths \cite{nielsen06}. In particular,
we are motivated here by the following questions: Can we geometrically
characterize the efficiency of quantum search schemes? Can we geometrically
quantify the effect of experimentally tunable parameters on the performance of
quantum search algorithms? Can we generate some fresh physical insight leading
to a (currently non-existing) geometric measure of thermal efficiency given
the fact that good quantum algorithms need to be both fast and
thermodynamically efficient?\textbf{ }More specifically, we wish to enhance in
this article our understanding of the time optimality of the original
Farhi-Gutmann quantum search Hamiltonian evolution \cite{farhi98} between the
generally nonorthogonal source and target states by gaining new insights with
the use of Riemannian geometric tools as originally proposed by Anandan and
Aharonov in Ref. \cite{anandan90}. In their work, Anandan and Aharonov
introduced the efficiency of a quantum evolution as $\eta\overset{\text{def}%
}{=}s_{0}/s$ with $s_{0}$ being the length of the geodesic path connecting the
initial and final quantum states of the system, while $s$ denotes the length
of the path generated by the actual Hamiltonian evolution. To our knowledge,
there are currently no explicit applications of $\eta$ in the literature,
which obscures both its physical meaning and its potential practical usefulness.

In this article, we present the first application of the geometric efficiency
of quantum evolutions using the Farhi-Gutmann quantum search Hamiltonian
\cite{farhi98} as an example. Given our previously mentioned considerations in
i), ii), and iii), we want to study the geometry of this quantum search
evolution between two orthogonal quantum states. In particular, we wish to
determine whether or not to a time optimal quantum search scheme, whose
analysis is based upon the concept of transition probability, there
corresponds a maximally efficient quantum search evolution achieving the ideal
unit efficiency value. Such a determination will be made in the scenario
wherein the output quantum state originating from the quantum search scheme
traces a shortest geodesic path connecting the suitably chosen initial and
final orthogonal states on the projective space (that is, the Bloch sphere $%
\mathbb{C}
P^{1}$ in the case of single-qubit quantum states) equipped with the
Fubini-Study metric. Furthermore, we wish to understand how deviations (see
Ref. \cite{carlophysica19}) from optimal quantum search schemes can be
described within the proposed Riemannian geometric framework and discuss any
physical insight that may arise from this theoretical description. Finally, we
wish to determine whether or not there exists any possibility of extending
this geometric characterization of quantum evolutions to the Riemannian
geometric study of thermal trajectories \cite{ruppeiner95a,quevedo07,brody98}.
The pursuit of such an extension is undertaken with the hope of proposing a
good geometric measure of thermal efficiency for thermodynamical processes of
interest in quantum information science \cite{castelvecchi97} by improving
upon our recent results in Refs. \cite{cafaropre18,cafaropre20}.

The layout of the remainder of this article is as follows. In Section II, we
briefly present the essential features of both the original and the modified
Farhi-Gutmann quantum search Hamiltonians. In particular, for each scheme, we
highlight both transition probabilities from the source to the target states
and the minimum search times yielding the maximum success probabilities. In
Section III, we introduce the essential features of the geometric structure of
quantum evolutions. More specifically, we describe the concept of a geodesic
line on the Bloch sphere and explain how to quantify a departure from a
geodesic evolution. In Section IV, we discuss a geometric measure of
efficiency for a quantum evolution together with its connection with a form of
time-energy uncertainty inequality to be satisfied during the physical
evolution at all times. In Section V, we study the geodesicity, the
efficiency, and the uncertainty inequality for both the original and modified
Farhi-Gutmann quantum search algorithms. We present our concluding remarks in
Section VI. Finally, some technical details can be found in Appendix A and
Appendix B.

\section{Quantum search Hamiltonians}

In this Section, we briefly discuss the main properties of both the original
\cite{farhi98} and the modified \cite{farhi98,carlophysica19} Farhi-Gutmann
quantum search algorithms. In particular, for each scheme, we emphasize both
transition probabilities from the source to the target states and the minimum
search times leading to the maximum success probabilities.

\subsection{The original scenario}

Quantum search algorithms, including Grover's original quantum search scheme
\cite{grover}, were originally proposed in a digital quantum computation
framework in terms of a discrete sequence of unitary logic gates. By contrast,
Farhi and Gutmann used an analog quantum computation setting to present an
analog version of Grover's original quantum search algorithm in which the
state of the quantum register undergoes a continuous time evolution under the
action of a conveniently selected driving Hamiltonian \cite{farhi98}. The
essential idea of the continuous time search algorithm proposed by Farhi and
Gutmann can be summarized as follows. Given an Hamiltonian acting on an
$N$-dimensional (with $N\overset{\text{def}}{=}2^{n}$) complex vector space
$\mathcal{H}_{2}^{n}$ with a single nonvanishing eigenvalue $E\neq0$ and all
others being zero, find the eigenvector $\left\vert w\right\rangle $ that has
eigenvalue equal to $E$. The search is completed when the quantum system is
known to be in the state $\left\vert w\right\rangle $. Working with
time-independent Hamiltonian evolutions, Farhi and Gutmann proved that their
algorithm required a minimum search time of the order $\sqrt{N}$, thus being
characterized by the same complexity as Grover's original quantum search
algorithm. The full original Farhi-Gutmann quantum search Hamiltonian is given
by \cite{farhi98},%
\begin{equation}
\mathrm{H}_{\text{FG}}\overset{\text{def}}{=}\mathrm{H}_{w}+\mathrm{H}%
_{d}=E\left\vert w\right\rangle \left\langle w\right\vert +E\left\vert
s\right\rangle \left\langle s\right\vert \text{,} \label{FGH}%
\end{equation}
with $\mathrm{H}_{w}\overset{\text{def}}{=}E\left\vert w\right\rangle
\left\langle w\right\vert $ and $\mathrm{H}_{d}\overset{\text{def}}%
{=}E\left\vert s\right\rangle \left\langle s\right\vert $ being the oracle and
driving Hamiltonians, respectively. The normalized states $\left\vert
s\right\rangle $ and $\left\vert w\right\rangle $ are the source (initial) and
target (final) states, respectively. The target state $\left\vert
w\right\rangle $ is a\ randomly chosen (unknown) state from the unit sphere in
$\mathcal{H}_{2}^{n}$, while the source state $\left\vert s\right\rangle $ is
some suitably selected normalized vector that does not depend on $\left\vert
w\right\rangle $. The source state $\left\vert s\right\rangle $ evolves
according to Schr\"{o}dinger's quantum mechanical evolution law \cite{peres95}%
,%
\begin{equation}
\left\vert s\right\rangle \mapsto\left\vert \psi\left(  t\right)
\right\rangle \overset{\text{def}}{=}e^{-\frac{i}{\hslash}\mathrm{H}%
_{\text{FG}}t}\left\vert s\right\rangle \text{.}%
\end{equation}
Moreover, without loss of generality, the quantum overlap $x\overset
{\text{def}}{=}\left\langle w|s\right\rangle \neq0$ can be taken to be real
and positive because any phase factor in the inner product between these two
states can be eventually incorporated in $\left\vert s\right\rangle $.
Moreover, given that it is sufficient to focus our attention to the
two-dimensional subspace of $\mathcal{H}_{2}^{n}$ spanned by $\left\vert
s\right\rangle $ and $\left\vert w\right\rangle $, it is convenient to
introduce the orthonormal basis\textbf{\ }$\left\{  \left\vert w\right\rangle
\text{, }\left\vert r\right\rangle \right\}  $ with $\left\vert r\right\rangle
\overset{\text{def}}{=}\left(  1-x^{2}\right)  ^{-1/2}\left(  \left\vert
s\right\rangle -x\left\vert w\right\rangle \right)  $ and $\left\vert
s\right\rangle \overset{\text{def}}{=}x\left\vert w\right\rangle
+\sqrt{1-x^{2}}\left\vert r\right\rangle $, respectively. Working with the
basis $\left\{  \left\vert w\right\rangle \text{, }\left\vert r\right\rangle
\right\}  $, it is possible to show that the probability $\mathcal{P}%
_{\text{FG}}\left(  t\right)  $ of finding the state $\left\vert
w\right\rangle $ at time time $t$ is given by \cite{farhi98},%
\begin{equation}
\mathcal{P}_{\text{FG}}\left(  t\right)  \overset{\text{def}}{=}\left\vert
\left\langle w|e^{-\frac{i}{\hslash}\mathrm{H}_{\text{FG}}t}|s\right\rangle
\right\vert ^{2}=\sin^{2}\left(  \frac{Ex}{\hslash}t\right)  +x^{2}\cos
^{2}\left(  \frac{Ex}{\hslash}t\right)  \text{.} \label{PFG}%
\end{equation}
\textbf{\ }In particular, the (smallest) instant $t_{\text{FG}}$ at which the
transition probability $\mathcal{P}_{\text{FG}}\left(  t\right)  $ assumes its
maximum value $\mathcal{P}_{\text{FG}}^{\max}=1$ is,%
\begin{equation}
t_{\text{FG}}\overset{\text{def}}{=}\frac{\pi\hslash}{2Ex}\text{.} \label{tgf}%
\end{equation}
When the target state $\left\vert w\right\rangle $ is assumed to be an element
of a set of mutually orthonormal quantum states $\left\{  \left\vert
a\right\rangle \right\}  $ with $1\leq a\leq N$ of $\mathcal{H}_{2}^{n}$, the
source state $\left\vert s\right\rangle $ can be conveniently chosen as an
equal superposition of the $N$ quantum states $\left\{  \left\vert
a\right\rangle \right\}  $. Then, $x=1/\sqrt{N}$ and from Eq. (\ref{tgf}) we
note that $t_{\text{FG}}\propto\sqrt{N}$. Thus, in analogy to Grover's search,
the Farhi-Gutmann algorithm requires a minimum search time of the order
$\sqrt{N}$. Additionally, by assuming that the target state is an unknown
element of a given orthonormal basis $\left\{  \left\vert a\right\rangle
\right\}  $ with $1\leq a\leq N$ of $\mathcal{H}_{2}^{n}$ that is produced
with absolute certainty, Farhi and Gutmann proved that their algorithm is
optimally short.

\subsection{The modified scenario}

Before considering their optimality proof, Farhi and Gutmann pointed out in
Ref. \cite{farhi98} that one may be driven by intuition to believe that by
using a different driving Hamiltonian $\mathrm{H}_{d}^{\prime}\overset
{\text{def}}{=}E^{\prime}\left\vert s\right\rangle \left\langle s\right\vert $
with $E^{\prime}\gg E$, one could shorten the search time by speeding up the
procedure for finding the target state $\left\vert w\right\rangle $. More
explicitly, consider the full modified Farhi-Gutmann quantum search
Hamiltonian given by \cite{farhi98,carlophysica19},%
\begin{equation}
\mathrm{H}_{\text{MFG}}\overset{\text{def}}{=}\mathrm{H}_{w}+\mathrm{H}%
_{d}^{\prime}=E\left\vert w\right\rangle \left\langle w\right\vert +E^{\prime
}\left\vert s\right\rangle \left\langle s\right\vert \text{,} \label{MFGH}%
\end{equation}
where $\mathrm{H}_{w}\overset{\text{def}}{=}E\left\vert w\right\rangle
\left\langle w\right\vert $ and $\mathrm{H}_{d}^{\prime}\overset{\text{def}%
}{=}E^{\prime}\left\vert s\right\rangle \left\langle s\right\vert $ with
$E^{\prime}\gg E$. Following the analysis performed in the original scenario,
it can be shown that the probability $\mathcal{P}_{\text{MFG}}\left(
t\right)  $ of finding the state $\left\vert w\right\rangle $ at time time $t$
is given by \cite{carlophysica19},%
\begin{equation}
\mathcal{P}_{\text{MFG}}\left(  t\right)  \overset{\text{def}}{=}\frac
{x^{2}\left(  E^{\prime}+E\right)  ^{2}}{4x^{2}E^{\prime}E+\left(  E^{\prime
}-E\right)  ^{2}}\sin^{2}\left(  \frac{1}{2\hslash}\sqrt{4x^{2}EE^{\prime
}+\left(  E^{\prime}-E\right)  ^{2}}t\right)  +x^{2}\cos^{2}\left(  \frac
{1}{2\hslash}\sqrt{4x^{2}EE^{\prime}+\left(  E^{\prime}-E\right)  ^{2}%
}t\right)  \text{.} \label{PMFG}%
\end{equation}
Observe that for $E=E^{\prime}$, we recover from Eq. (\ref{PMFG}) the
expression of $\mathcal{P}_{\text{FG}}\left(  t\right)  $ in Eq. (\ref{PFG}).
Moreover, the (smallest) instant $t_{\text{MFG}}$ at which the transition
probability $\mathcal{P}_{\text{MFG}}\left(  t\right)  $ assumes its maximum
value $\mathcal{P}_{\text{MFG}}^{\max}=\left[  x^{2}\left(  E^{\prime
}+E\right)  ^{2}\right]  /\left[  4x^{2}E^{\prime}E+\left(  E^{\prime
}-E\right)  ^{2}\right]  <1$ is,%
\begin{equation}
t_{\text{MFG}}\overset{\text{def}}{=}\frac{\pi\hslash}{\sqrt{4x^{2}E^{\prime
}E+\left(  E^{\prime}-E\right)  ^{2}}}\text{.} \label{tmgf}%
\end{equation}
As expected, when $E^{\prime}=E$, $t_{\text{MFG}}$ in Eq. (\ref{tmgf}) reduces
to $t_{\text{FG}}$ in Eq. (\ref{tgf}). Clearly, by\textbf{ }comparing the
transition probabilities in Eqs. (\ref{PFG}) and (\ref{PMFG}), we are able
to\textbf{ }conclude in a transparent manner that using a modified driving
Hamiltonian $\mathrm{H}_{d}^{\prime}\overset{\text{def}}{=}E^{\prime
}\left\vert s\right\rangle \left\langle s\right\vert $ with $E^{\prime}\gg E$
does not speed up the procedure for producing the target state $\left\vert
w\right\rangle $ with certainty. Indeed, although $t_{\text{MFG}}$ in Eq.
(\ref{tmgf}) is smaller than $t_{\text{FG}}$ in Eq. (\ref{tgf}), we note that
$\mathcal{P}_{\text{MFG}}^{\max}<\mathcal{P}_{\text{FG}}^{\max}=1$. Therefore,
while the Hamiltonian $\mathrm{H}_{\text{MFG}}$ may have some merit in nearly
optimal quantum search schemes as pointed out in Ref. \cite{carlophysica19},
it appears to be less \textquotedblleft efficient\textquotedblright\ than
$\mathrm{H}_{\text{FG}}$ and consequently, does not lead to any advantage in
the context of quantum search with certainty as one may have thought from a
classically intuitive point of view. Despite the Farhi-Gutmann formal
optimality proof and the Cafaro-Alsing brute force transition probability
analysis, it remains interesting to consider whether or not the different
\textquotedblleft efficiency\textquotedblright\ of the quantum search schemes
specified by the Hamiltonians $\mathrm{H}_{\text{FG}}$ and $\mathrm{H}%
_{\text{MFG}}$ can be understood in neat geometric terms that might be closer
to our intuition. Motivated by this main thought, we propose in what follows a
geometric perspective on the efficiency of these two analog quantum search schemes.

\section{Geodesics in ray space}

In this Section, we introduce basic geometric concepts of quantum evolutions
with emphasis on the unitary Schr\"{o}dinger evolution.

Let $\mathcal{H}_{2}^{n}$ denote an $N\overset{\text{def}}{=}2^{n}%
$-dimensional complex Hilbert space of $n$-qubit (normalized) quantum states
$\left\{  \left\vert \psi\right\rangle \right\}  $. Since the global phase of
a vector state is not observable, a physical state is represented by a
so-called ray of the Hilbert space. The set of rays of $\mathcal{H}_{2}^{n}$
is called the (complex) projective Hilbert space $%
\mathbb{C}
P^{N-1}$. Formally speaking, $%
\mathbb{C}
P^{N-1}$ is the quotient set of $\mathcal{H}_{2}^{n}$ by the equivalence
relation $\left\vert \psi\right\rangle \sim e^{i\beta}\left\vert
\psi\right\rangle $ with $\beta\in%
\mathbb{R}
$. The space $%
\mathbb{C}
P^{N-1}$ can be equipped with a mathematically correct and physically
meaningful metric structure. Indeed, consider a family $\left\{  \left\vert
\psi\left(  \xi\right)  \right\rangle \right\}  $ of normalized quantum states
of $\mathcal{H}_{2}^{n}$ that smoothly depend on an $m$-dimensional parameter
$\xi\overset{\text{def}}{=}\left(  \xi^{1}\text{,..., }\xi^{m}\right)  \in%
\mathbb{R}
^{m}$. Then, the ordinary Hermitian scalar product on $\mathcal{H}_{2}^{n}$
induces a metric tensor $g_{ab}\left(  \xi\right)  $ with $1\leq a$, $b\leq m$
on the manifold of quantum states defined as \cite{provost80},%
\begin{equation}
g_{ab}\left(  \xi\right)  \overset{\text{def}}{=}4\operatorname{Re}\left[
\left\langle \partial_{a}\psi\left(  \xi\right)  |\partial_{b}\psi\left(
\xi\right)  \right\rangle -\left\langle \partial_{a}\psi\left(  \xi\right)
|\psi\left(  \xi\right)  \right\rangle \left\langle \psi\left(  \xi\right)
|\partial_{b}\psi\left(  \xi\right)  \right\rangle \right]  \text{,}%
\label{fubini}%
\end{equation}
with $\partial_{a}\overset{\text{def}}{=}\partial/\partial\xi^{a}$. The
quantity $g_{ab}\left(  \xi\right)  $ in Eq. (\ref{fubini}) is the so-called
Fubini-Study metric tensor. In particular, we note that the metric is positive
definite as is evident by considering the distance element $ds_{\text{FS}}%
^{2}$ between two nearby points with associated vector states $\left\vert
\psi\left(  \xi+d\xi\right)  \right\rangle $ and $\left\vert \psi\left(
\xi\right)  \right\rangle $ \cite{brau94},%
\begin{equation}
ds_{\text{FS}}^{2}\overset{\text{def}}{=}g_{ab}\left(  \xi\right)  d\xi
^{a}d\xi^{b}=4\left[  \left\langle d\psi|d\psi\right\rangle -\left\vert
\left\langle \psi|d\psi\right\rangle \right\vert ^{2}\right]  \text{,}%
\label{distance1}%
\end{equation}
where $\left\vert d\psi\right\rangle \overset{\text{def}}{=}\left\vert
\psi\left(  \xi+d\xi\right)  \right\rangle -\left\vert \psi\left(  \xi\right)
\right\rangle $. The distance element in Eq. (\ref{distance1}) leads naturally
to the concept of geodesic paths in $%
\mathbb{C}
P^{N-1}$. Indeed, by using variational calculus arguments, geodesic paths in $%
\mathbb{C}
P^{N-1}$ can be obtained by minimizing the distance integral \textrm{S
}\cite{grigorenko92},%
\begin{equation}
\mathrm{S}\overset{\text{def}}{=}\int ds_{\text{FS}}=2\int\left[  \left\langle
d\psi|d\psi\right\rangle -\left\vert \left\langle \psi|d\psi\right\rangle
\right\vert ^{2}\right]  ^{1/2}=\int\mathcal{L}d\tau\text{,}%
\end{equation}
with $\mathcal{L}\overset{\text{def}}{=}2\left[  \left\langle \dot{\psi}%
|\dot{\psi}\right\rangle -\left\vert \left\langle \psi|\dot{\psi}\right\rangle
\right\vert ^{2}\right]  ^{1/2}$, $\left\vert \dot{\psi}\right\rangle
\overset{\text{def}}{=}\partial_{\tau}\left\vert \psi\right\rangle $, and
$\tau$ being a parameter along the curve $\gamma\left(  \tau\right)
:\tau\mapsto\left\vert \psi\left(  \tau\right)  \right\rangle $ that we assume
to be equal to the natural parameter $s_{\text{FS}}=s$. We recall that if
$\left\vert \psi\left(  s\right)  \right\rangle $ is a geodesic then the
phase-transformed vector $\left\vert \bar{\psi}\left(  s\right)  \right\rangle
\overset{\text{def}}{=}e^{i\beta\left(  s\right)  }\left\vert \psi\left(
s\right)  \right\rangle $ with arbitrary $\beta\left(  s\right)  $ is also a
geodesic \cite{grigorenko92}. In particular, by conveniently choosing
$\beta\left(  s\right)  $ such that $\left\langle \bar{\psi}\left(  s\right)
|\bar{\psi}^{\prime}\left(  s\right)  \right\rangle =0$ with $\left\vert
\bar{\psi}^{\prime}\left(  s\right)  \right\rangle \overset{\text{def}}%
{=}\partial_{s}\left\vert \bar{\psi}\left(  s\right)  \right\rangle $ (that
is, $\left\vert \bar{\psi}\left(  s\right)  \right\rangle $ is the horizontal
lift of $\left\vert \psi\left(  s\right)  \right\rangle $ satisfying the
parallel transport rule), it can be shown after some straightforward but
tedious variational calculus computations that a geodesic $\left\vert
\bar{\psi}\left(  s\right)  \right\rangle $ satisfies a simple harmonic
oscillator equation \cite{mukunda93,pati94},%
\begin{equation}
\left\vert \bar{\psi}^{\prime\prime}\left(  s\right)  \right\rangle
+\left\vert \bar{\psi}\left(  s\right)  \right\rangle =0\text{.}\label{geoeq}%
\end{equation}
Assuming $\left\langle \bar{\psi}\left(  0\right)  |\bar{\psi}\left(
0\right)  \right\rangle =1$, $\left\langle \bar{\psi}\left(  0\right)
|\bar{\psi}^{\prime}\left(  0\right)  \right\rangle =0$, and $\left\langle
\bar{\psi}^{\prime}\left(  0\right)  |\bar{\psi}^{\prime}\left(  0\right)
\right\rangle =\omega^{2}$ with $\omega$ being\textbf{ }a constant in $%
\mathbb{R}
$, the solution of Eq. (\ref{geoeq}) can be written as,%
\begin{equation}
\left\vert \bar{\psi}\left(  s\right)  \right\rangle =\cos\left(  \omega
s\right)  \left\vert \bar{\psi}\left(  0\right)  \right\rangle +\frac
{\sin\left(  \omega s\right)  }{\omega}\left\vert \bar{\psi}^{\prime}\left(
0\right)  \right\rangle \text{.}\label{geo2}%
\end{equation}
Eq. (\ref{geo2}) represents the most general geodesic in horizontal and
affinely parameterized form in $%
\mathbb{C}
P^{N-1}$ \cite{mukunda93}. More generally, it can be shown that any two
arbitrary unit vectors $\left\vert \psi_{A}\right\rangle $ and $\left\vert
\psi_{B}\right\rangle $ in the projective Hilbert space can be connected by a
geodesic line $\left\vert \psi_{\text{geo}}\left(  \lambda\right)
\right\rangle $ parametrized by a real parameter $0\leq\lambda\leq1$
\cite{mukunda93,laba17},%
\begin{equation}
\left\vert \psi_{\text{geo}}\left(  \lambda\right)  \right\rangle
\overset{\text{def}}{=}\frac{\left(  1-\lambda\right)  \left\vert \psi
_{A}\right\rangle +e^{i\phi}\lambda\left\vert \psi_{B}\right\rangle }%
{\sqrt{1-2\lambda\left(  1-\lambda\right)  \left[  1-\left\vert \left\langle
\psi_{B}|\psi_{A}\right\rangle \right\vert \right]  }}\text{,}\label{geoline}%
\end{equation}
where $\left\vert \psi_{\text{geo}}\left(  0\right)  \right\rangle =\left\vert
\psi_{A}\right\rangle $, $\left\vert \psi_{\text{geo}}\left(  1\right)
\right\rangle =\left\vert \psi_{B}\right\rangle $, and $\phi\in%
\mathbb{R}
$ with $\left\langle \psi_{B}|\psi_{A}\right\rangle =\left\vert \left\langle
\psi_{B}|\psi_{A}\right\rangle \right\vert e^{i\phi}$. For the sake of
completeness, we emphasize that a simple explicit way to check that
$\left\vert \psi_{\text{geo}}\left(  \lambda\right)  \right\rangle $ does
indeed represent a geodesic line\textbf{ }is to show that the length of the
curve connecting $\left\vert \psi_{A}\right\rangle $ and $\left\vert \psi
_{B}\right\rangle $ measured with the Fubini-Study metric equals the minimal
possible length of the curve on the Bloch sphere connecting these two states.
Upon recasting Eq. (\ref{geoline}) as,%
\begin{equation}
\left\vert \psi_{\text{geo}}\left(  \theta\right)  \right\rangle
\overset{\text{def}}{=}\frac{\cos\left(  \frac{\theta}{2}\right)  \left\vert
\psi_{A}\right\rangle +e^{i\phi}\sin\left(  \frac{\theta}{2}\right)
\left\vert \psi_{B}\right\rangle }{\sqrt{1+\sin\left(  \theta\right)
\left\vert \left\langle \psi_{B}|\psi_{A}\right\rangle \right\vert }}%
\text{,}\label{geoline2}%
\end{equation}
with $\lambda=\lambda\left(  \theta\right)  \overset{\text{def}}{=}\tan\left(
\theta/2\right)  /\left[  1+\tan\left(  \theta/2\right)  \right]  $ being a
strictly monotonic function of $\theta$ where $0\leq\theta\leq\pi$, it can be
shown that the length $s\overset{\text{def}}{=}\int_{0}^{\pi}\sqrt
{ds_{\text{FS}}^{2}}$ of this curve equals $2\cos^{-1}\left[  \left\vert
\left\langle \psi_{B}|\psi_{A}\right\rangle \right\vert \right]  $. This, in
turn, coincides with the Wootters distance or, equivalently, the angle between
the two states $\left\vert \psi_{A}\right\rangle $ and $\left\vert \psi
_{B}\right\rangle $ \cite{wootters81}. Thus, $\left\vert \psi_{\text{geo}%
}\left(  \lambda\right)  \right\rangle $ and $\left\vert \psi_{\text{geo}%
}\left(  \theta\right)  \right\rangle $ in Eqs. (\ref{geoline}) and
(\ref{geoline2}) respectively, are indeed geodesic arcs. We point out that in
Eqs. (\ref{geoline}) and (\ref{geoline2}), it is assumed that $\left\vert
\psi_{A}\right\rangle $ and $\left\vert \psi_{B}\right\rangle $ are
nonorthogonal. When $\left\vert \psi_{A}\right\rangle \perp\left\vert \psi
_{B}\right\rangle $, geodesic lines can be obtained from Eqs. (\ref{geoline})
and (\ref{geoline2}) by taking $\phi=0$ and, clearly, $\left\langle \psi
_{B}|\psi_{A}\right\rangle =0$. One way to determine whether or not
Schr\"{o}dinger's solution $\left\vert \psi\left(  t\right)  \right\rangle $
specifies a geodesic path is to verify that the geodesic curvature of its
corresponding dynamical trajectory on the Bloch sphere is identically zero.
Alternatively, a more convenient approach available to us is to check whether
there exists a reversible mapping $\lambda:\left(  0\text{, }t_{\ast}\right)
\ni t\mapsto\lambda\left(  t\right)  \in\left(  0\text{, }1\right)  $ with
$\lambda(0)=0$ and $\lambda\left(  t_{\ast}\right)  =1$ such that the distance
$d^{2}\left(  t\text{, }\lambda\right)  $ between $\left\vert \psi\left(
t\right)  \right\rangle $ and $\left\vert \psi_{\text{geo}}\left(
\lambda\right)  \right\rangle $ in Eq. (\ref{geoline2}),%
\begin{equation}
d^{2}\left(  t\text{, }\lambda\right)  \overset{\text{def}}{=}4\left[
1-\left\vert \left\langle \psi\left(  t\right)  |\psi_{\text{geo}}\left(
\lambda\right)  \right\rangle \right\vert ^{2}\right]  \text{,}%
\label{distanza}%
\end{equation}
is identically zero. As a final remark of geometric flavor, we point out that
horizontal affinely parametrized geodesics on the Bloch sphere are great
circles traced by state vectors as in Eq. (\ref{geo2})
\cite{provost80,bhandari88,carol}. 

The quantum material presented in Section II together with the geometric
material covered in Section III will be helpful in putting the concepts of
efficiency and uncertainty of search evolutions to be introduced in the next
Section in the proper geometric formulation of quantum evolutions as
originally proposed by Anandan and Aharonov in Ref. \cite{anandan90}.

\section{Efficiency and uncertainty of quantum evolutions}

In this Section, following the work by Anandan and Aharonov in Ref.
\cite{anandan90}, we discuss a geometric measure of efficiency for a quantum
search evolution together with its connection to a form of time-energy
uncertainty inequality, with the latter being fulfilled at all times during
the physical evolution of the system under consideration.

\subsection{Efficiency}

Consider a quantum mechanical evolution of a state vector $\left\vert
\psi\left(  t\right)  \right\rangle $ described by the Schr\"{o}dinger
equation,%
\begin{equation}
i\hslash\partial_{t}\left\vert \psi\left(  t\right)  \right\rangle
=\mathrm{H}\left(  t\right)  \left\vert \psi\left(  t\right)  \right\rangle
\text{,} \label{SS}%
\end{equation}
with $0\leq t\leq t_{\ast}$. Following the work by Anandan and Aharonov, a
geometric measure of efficiency for such a quantum evolution can be formally
defined as \cite{anandan90},%
\begin{equation}
\eta\overset{\text{def}}{=}1-\frac{\Delta s}{s}=\frac{2\cos^{-1}\left[
\left\vert \left\langle \psi\left(  0\right)  |\psi\left(  t_{\ast}\right)
\right\rangle \right\vert \right]  }{2\int_{0}^{t_{\ast}}\frac{\Delta E\left(
t^{\prime}\right)  }{\hslash}dt^{\prime}}\text{,} \label{efficiency}%
\end{equation}
where $\Delta s\overset{\text{def}}{=}s-s_{0}$, $s_{0}$ denotes the distance
along the shortest geodesic path joining the distinct initial $\left\vert
\psi\left(  0\right)  \right\rangle $ and final $\left\vert \psi\left(
t_{\ast}\right)  \right\rangle $ states on the projective Hilbert space $%
\mathbb{C}
P^{N-1}$ and finally, $s$ is the distance along the actual dynamical
trajectory traced by the state vector $\left\vert \psi\left(  t\right)
\right\rangle $ with $0\leq t\leq t_{\ast}$. Observe that the numerator
in\ Eq. (\ref{efficiency}) is the angle between the state vectors $\left\vert
\psi\left(  0\right)  \right\rangle $ and $\left\vert \psi\left(  t_{\ast
}\right)  \right\rangle $ and equals the Wootters distance
$ds_{\text{Wootters}}$ \cite{wootters81},%
\begin{equation}
ds_{\text{Wootters}}\left(  \left\vert \psi\left(  t_{1}\right)  \right\rangle
\text{, }\left\vert \psi\left(  t_{2}\right)  \right\rangle \right)
\overset{\text{def}}{=}2\cos^{-1}\left[  \left\vert \left\langle \psi\left(
t_{1}\right)  |\psi\left(  t_{2}\right)  \right\rangle \right\vert \right]
\text{.}%
\end{equation}
The denominator in Eq. (\ref{efficiency}) instead, is the integral of the
infinitesimal distance $ds$ along the evolution curve (that is, the actual
dynamical trajectory) in the projective Hilbert space \cite{anandan90},%
\begin{equation}
ds\overset{\text{def}}{=}2\frac{\Delta E\left(  t\right)  }{\hslash}dt\text{,}
\label{distance}%
\end{equation}
with $\Delta E$ being the square root of the dispersion (or equivalently, the
variance) of the Hamiltonian operator $\mathrm{H}\left(  t\right)  $,%
\begin{equation}
\Delta E\left(  t\right)  \overset{\text{def}}{=}\left[  \left\langle
\psi|\mathrm{H}^{2}\left(  t\right)  |\psi\right\rangle -\left\langle
\psi|\mathrm{H}\left(  t\right)  |\psi\right\rangle ^{2}\right]
^{1/2}\text{.} \label{dispersion}%
\end{equation}
Interestingly, Anandan and Aharonov showed that the infinitesimal distance
$ds$ in Eq. (\ref{distance}) is related to the Fubini-Study infinitesimal
distance $ds_{\text{Fubini-Study}}$ by the following relation,%
\begin{equation}
ds_{\text{Fubini-Study}}^{2}\left(  \left\vert \psi\left(  t\right)
\right\rangle \text{, }\left\vert \psi\left(  t+dt\right)  \right\rangle
\right)  \overset{\text{def}}{=}4\left[  1-\left\vert \left\langle \psi\left(
t\right)  |\psi\left(  t+dt\right)  \right\rangle \right\vert ^{2}\right]
=4\frac{\Delta E^{2}\left(  t\right)  }{\hslash^{2}}dt^{2}+\mathcal{O}\left(
dt^{3}\right)  \text{,} \label{relation}%
\end{equation}
with $\mathcal{O}\left(  dt^{3}\right)  $ denoting an infinitesimal quantity
equal or higher than $dt^{3}$. From Eqs. (\ref{distance}) and (\ref{relation}%
), it follows that $s$ is proportional to the time integral of the uncertainty
in energy $\Delta E$ of the system and represents the distance along the
quantum evolution of the physical system in $%
\mathbb{C}
P^{N-1}$ as measured by the Fubini-Study metric. We point out that when the
actual dynamical curve coincides with the shortest geodesic path connecting
the initial and final states, $\Delta s$ equals zero and the efficiency $\eta$
in Eq. (\ref{efficiency}) becomes one. Clearly, the shortest possible distance
between two orthogonal quantum states on $%
\mathbb{C}
P^{N-1}$ is $\pi$ while, in general $s\geq\pi$ for such a pair of orthogonal
pure states. These considerations will become especially useful in Section V.
For the interested readers, we confine a brief discussion on possible
generalizations of $\eta$ in\ Eq. (\ref{efficiency}) to geometric evolutions
of mixed quantum states not limited to temporal unitary propagators in
Appendix A. In the next subsection, we elaborate on the concept of uncertainty
of a quantum search evolution.

\subsection{Uncertainty}

In quantum theory \cite{peres95}, the standard quantum mechanical uncertainty
relation given by%
\begin{equation}
\Delta x\Delta p\geq\hslash/2\text{,} \label{heisenberg}%
\end{equation}
reflects the intrinsic randomness of the outcomes of quantum
experiments.\ Specifically, if one repeats many times the same state
preparation scheme and then measures the operators $x$ or $p$, the variety of
observations recorded for $x$ and $p$ are characterized by standard deviations
$\Delta x$ and $\Delta p$ whose product $\Delta x\Delta p$ is greater than
$\hslash/2$. In particular, Gaussian wave packets (for instance, the ground
state of a shifted harmonic oscillator) are specified by a minimum
position-momentum uncertainty with $\Delta x\Delta p=\hslash/2$.

In the geometry of quantum evolutions, there exists an analog of Eq.
(\ref{heisenberg}) on the one hand, while on the other, Gaussian wave packets
are replaced by geodesic paths in the projective Hilbert space. Indeed,
consider the time-averaged uncertainty in energy $\left\langle \Delta
E\right\rangle $ during a time interval $\Delta t_{\perp}$ defined as
\cite{anandan90},%
\begin{equation}
\left\langle \Delta E\right\rangle \overset{\text{def}}{=}\frac{1}{\Delta
t_{\perp}}\int_{0}^{\Delta t_{\perp}}E\left(  t^{\prime}\right)  dt^{\prime
}\text{.} \label{energy}%
\end{equation}
The quantity $\Delta t_{\perp}$ in Eq. (\ref{energy}) represents the
orthogonalization time, that is, the time interval during which the system
passes from an initial state $\left\vert \psi_{A}\right\rangle \overset
{\text{def}}{=}\left\vert \psi\left(  0\right)  \right\rangle $ to a final
state $\left\vert \psi_{B}\right\rangle \overset{\text{def}}{=}\left\vert
\psi\left(  \Delta t_{\perp}\right)  \right\rangle $ where $\left\langle
\psi_{B}|\psi_{A}\right\rangle =\delta_{AB}$. Using Eqs. (\ref{distance}) and
(\ref{energy}) and recalling that the shortest possible distance between two
orthogonal quantum states on $%
\mathbb{C}
P^{N-1}$ is $\pi$, we get%
\begin{equation}
\Delta\overset{\text{def}}{=}\left\langle \Delta E\right\rangle \Delta
t_{\perp}\geq h/4\text{.} \label{uncertainty}%
\end{equation}
In particular, it is\textbf{ }only when the quantum evolution is a geodesic
evolution that the equality in Eq. (\ref{uncertainty}) holds. Thus, just as
Gaussian wave packets are minimum position-momentum uncertainty wave packets,
geodesic paths are minimum time-averaged energy uncertainty trajectories. In
summary, unit efficiency $\eta=1$ is achieved when a quantum evolution has
minimum uncertainty $\left\langle \Delta E\right\rangle \Delta t_{\perp}=h/4$.
This, in turn, happens only if the physical systems moves along a geodesic
path in $%
\mathbb{C}
P^{N-1}$. Interestingly, the Anandan-Aharonov time-energy uncertainty relation
in Eq. (\ref{uncertainty}) can be linked to the statistical speed of
evolution\textbf{ }$ds_{\text{FS}}/dt$\textbf{\ }of the physical system
with\textbf{ }$ds_{\text{FS}}^{2}$\textbf{\ }being the Fubini-Study
infinitesimal line element squared. Specifically, since\textbf{ }%
$ds_{\text{FS}}/dt=\Delta E\left(  t\right)  /\hslash$\textbf{,} the physical
system moves expeditiously wherever the uncertainty in energy is large.

The concept of geodesic line mentioned in Section III together with the
concepts of efficiency and uncertainty presented in this Section will be used
in the next Section in order to geometrically analyze the quantum search
evolutions described in Section II.

\section{Geodesicity, efficiency, and uncertainty of quantum search
evolutions}

In this Section, we aim to study the geodesicity condition $d^{2}\left(
t\text{, }\lambda\right)  =0$ with $d^{2}\left(  t\text{, }\lambda\right)  $
in\ Eq. (\ref{distanza}), the efficiency in Eq. (\ref{efficiency}), and the
uncertainty inequality in Eq. (\ref{uncertainty}) for both the original and
modified Farhi-Gutmann quantum search algorithms presented in Section II.

\subsection{The original scenario}

Considering the original Farhi-Gutmann scenario, the state vector $\left\vert
\psi\left(  t\right)  \right\rangle $ that solves the Schr\"{o}dinger
evolution relation in Eq. (\ref{SS}) with $\mathrm{H}=\mathrm{H}_{\text{FG}}$
in Eq. (\ref{FGH}) such that $\left\vert \psi_{A}\right\rangle =\left\vert
\psi\left(  0\right)  \right\rangle $, $\left\vert \psi_{B}\right\rangle
=\left\vert \psi\left(  t_{\ast}\right)  \right\rangle $, $\left\langle
\psi_{B}|\psi_{A}\right\rangle \overset{x\rightarrow0}{\longrightarrow}%
\delta_{AB}$ with $x\overset{\text{def}}{=}\left\langle w|s\right\rangle $,
and $t_{\ast}=\Delta t_{\perp}=t_{\text{FG}}$ with $t_{\text{FG}}$ defined in
Eq. (\ref{tgf}) is given by,%
\begin{equation}
\left\vert \psi\left(  t\right)  \right\rangle =\frac{1}{\sqrt{2}}%
\frac{e^{-\frac{i}{\hslash}Et}}{\sqrt{1-\sqrt{1-x^{2}}}}\left(
\begin{array}
[c]{c}%
\left(  1-\sqrt{1-x^{2}}\right)  \cos\left(  \frac{Ex}{\hslash}t\right)
-ix\sin\left(  \frac{Ex}{\hslash}t\right)  \\
x\cos\left(  \frac{Ex}{\hslash}t\right)  +i\left(  1-\sqrt{1-x^{2}}\right)
\sin\left(  \frac{Ex}{\hslash}t\right)
\end{array}
\right)  \text{.}\label{originals}%
\end{equation}
Substituting Eqs. (\ref{geoline}) and (\ref{originals}) into $d^{2}\left(
t\text{, }\lambda\right)  $ in Eq. (\ref{distanza}), we obtain%
\begin{equation}
d_{\text{FG}}^{2}\left(  t\text{, }\lambda\right)  =4\left\{  1-\left[
\frac{\left(  1-\lambda\right)  ^{2}\cos^{2}\left(  \frac{Ex}{\hslash
}t\right)  }{1-2\lambda\left(  1-\lambda\right)  }+\frac{\lambda^{2}\sin
^{2}\left(  \frac{Ex}{\hslash}t\right)  }{1-2\lambda\left(  1-\lambda\right)
}+\frac{\lambda\left(  1-\lambda\right)  \sin\left(  \frac{2Ex}{\hslash
}t\right)  \cos\left(  \frac{\pi}{2x}\right)  }{1-2\lambda\left(
1-\lambda\right)  }\right]  \right\}  \text{.}\label{lt0}%
\end{equation}
Finally, we impose $d^{2}\left(  t\text{, }\lambda\right)  $ equal to zero so
as\textbf{ }to find possible roots $\left\{  \lambda\left(  t\right)
\right\}  $. Then, in order to obtain a well-defined reversible mapping
$\lambda:\left(  0\text{, }t_{\ast}\right)  \ni t\mapsto\lambda\left(
t\right)  \in\left(  0\text{, }1\right)  $ with $\lambda(0)=0$ and
$\lambda\left(  t_{\ast}\right)  =1$, we find it is necessary to have
$t_{\ast}=t_{\text{FG}}(x)\overset{\text{def}}{=}\pi\hslash/(2Ex)$ with
$x\in\left(  0\text{, }1\right)  $ such that%
\begin{equation}
\frac{\pi}{2x}=2n\pi\text{,}\label{solvable}%
\end{equation}
with $n\in%
\mathbb{N}
$. Finally, given that Eq. (\ref{solvable}) is clearly solvable, we find that
a suitable reversible mapping $\lambda\left(  t\right)  $ is given by
\begin{equation}
\lambda\left(  t\right)  \overset{\text{def}}{=}\frac{\sin^{2}\left(  \frac
{E}{4\hslash}t\right)  +\frac{1}{2}\sin\left(  \frac{E}{2\hslash}t\right)
}{1+\sin\left(  \frac{E}{2\hslash}t\right)  }\text{.}\label{lt}%
\end{equation}
Indeed, it is straightforward to check that $d_{\text{FG}}^{2}\left(  t\text{,
}\lambda\right)  =0$ by substituting Eqs. (\ref{solvable}) and (\ref{lt}) into
Eq. (\ref{lt0}). Interestingly, observe that $\lambda\left(  t\right)  $ in
Eq. (\ref{lt}) is a strictly monotonic increasing function of $t$ with $0\leq
t\leq t_{\ast}$ and is such that $\lambda\left(  t_{\ast}/2\right)  =1/2$. We
remark that we have shown that the Farhi-Gutmann Hamiltonian evolution
trajectory between the two orthogonal quantum states $\left\vert \psi
_{A}\right\rangle $ and $\left\vert \psi_{B}\right\rangle $ is formally a
geodesic in the limiting scenario in which the quantum overlap $x$ approaches
zero, that is, the duration of the evolution $t_{\text{FG}}$ approaches
infinity (long-time-limit). This limit requires formally selecting a very
large value of $n$ in Eq. (\ref{solvable}) when defining our reversible
mapping $\lambda=\lambda\left(  t\right)  $. This requirement is physically
consistent with the fact that $t_{\text{FG}}$ , being inversely proportional
to $x$, tends to diverge when the Farhi-Gutmann Hamiltonian evolution occurs
between nearly orthogonal source and target quantum states. It is in this
regime that we conduct our analysis of Farhi-Gutmann and modified
Farhi-Gutmann trajectories in this paper. For additional comments on this
point, we refer to Appendix B.\textbf{ }Finally, by substituting Eq.
(\ref{originals}) into the efficiency $\eta$ given in Eq. (\ref{efficiency})
and the time-energy uncertainty inequality presented in Eq. (\ref{uncertainty}%
), we obtain%
\begin{equation}
\eta_{\text{FG}}=1\text{, and }\Delta_{\text{FG}}\overset{\text{def}}%
{=}\left[  \left\langle \Delta E\right\rangle \Delta t_{\perp}\right]
_{_{\text{FG}}}=h/4\text{,}\label{unit}%
\end{equation}
respectively. Thus by investigating the geometry of the original Farhi-Gutmann
quantum evolution, we are able to conclude that it describes a geodesic motion
on the Bloch sphere specified by unit efficiency $\eta_{\text{FG}}$ and
a\textbf{ }minimum uncertainty $\Delta_{\text{FG}}$ that reaches the minimum
achievable value of $h/4$.

\subsection{The modified scenario}

Within the context of the modified Farhi-Gutmann scenario, the state vector
$\left\vert \psi\left(  t\right)  \right\rangle $ that solves the
Schr\"{o}dinger evolution relation in Eq. (\ref{SS}) with $\mathrm{H}%
=\mathrm{H}_{\text{MFG}}$ in Eq. (\ref{MFGH}) such that $\left\vert \psi
_{A}\right\rangle =\left\vert \psi\left(  0\right)  \right\rangle $,
$\left\vert \psi_{B}\right\rangle =\left\vert \psi\left(  t_{\ast}\right)
\right\rangle $, $\left\langle \psi_{B}|\psi_{A}\right\rangle \overset
{x\rightarrow0}{\longrightarrow}\delta_{AB}$ with $x\overset{\text{def}}%
{=}\left\langle w|s\right\rangle $, and $t_{\ast}=\Delta t_{\perp
}=t_{\text{MFG}}$ with $t_{\text{MFG}}$ defined in Eq. (\ref{tmgf}) is given
by,%
\begin{equation}
\left\vert \psi\left(  t\right)  \right\rangle =e^{-\frac{i}{\hslash}%
\frac{E^{\prime}+E}{2}t}\left(
\begin{array}
[c]{cc}%
\cos\left(  \frac{\lambda}{\hslash}t\right)  +i\frac{A+B}{A-B}\sin\left(
\frac{\lambda}{\hslash}t\right)   & -2i\frac{AB}{A-B}\sin\left(  \frac
{\lambda}{\hslash}t\right)  \\
\frac{2i}{A-B}\sin\left(  \frac{\lambda}{\hslash}t\right)   & \cos\left(
\frac{\lambda}{\hslash}t\right)  -i\frac{A+B}{A-B}\sin\left(  \frac{\lambda
}{\hslash}t\right)
\end{array}
\right)  \left\vert \psi\left(  0\right)  \right\rangle \text{.}\label{gpsi}%
\end{equation}
The initial state $\left\vert \psi\left(  0\right)  \right\rangle $ in Eq.
(\ref{gpsi}) is defined as,%
\begin{equation}
\left\vert \psi\left(  0\right)  \right\rangle \overset{\text{def}}{=}\left(
\begin{array}
[c]{c}%
\frac{\sqrt{\left(  1-AB\right)  ^{2}+\left(  A+B\right)  ^{2}}-\left(
1-AB\right)  }{\sqrt{\left(  A+B\right)  ^{2}+\left(  \sqrt{\left(
1-AB\right)  ^{2}+\left(  A+B\right)  ^{2}}-\left(  1-AB\right)  \right)
^{2}}}\\
\frac{A+B}{\sqrt{\left(  A+B\right)  ^{2}+\left(  \sqrt{\left(  1-AB\right)
^{2}+\left(  A+B\right)  ^{2}}-\left(  1-AB\right)  \right)  ^{2}}}%
\end{array}
\right)  \text{,}%
\end{equation}
where the quantities $A=A\left(  x\text{, }E^{\prime}\text{, }E\right)  $ and
$B=B\left(  x\text{, }E^{\prime}\text{, }E\right)  $ are explicitly given by%
\begin{equation}
A\left(  x\text{, }E^{\prime}\text{, }E\right)  \overset{\text{def}}{=}%
\frac{1}{2xE^{\prime}\sqrt{1-x^{2}}}\left[  E-E^{\prime}+2x^{2}E^{\prime
}-\sqrt{4x^{2}EE^{\prime}+\left(  E^{\prime}-E\right)  ^{2}}\right]  \text{, }%
\end{equation}
and,%
\begin{equation}
B\left(  x\text{, }E^{\prime}\text{, }E\right)  \overset{\text{def}}{=}%
\frac{1}{2xE^{\prime}\sqrt{1-x^{2}}}\left[  E-E^{\prime}+2x^{2}E^{\prime
}+\sqrt{4x^{2}EE^{\prime}+\left(  E^{\prime}-E\right)  ^{2}}\right]  \text{,}%
\end{equation}
respectively. Finally, the quantity $\lambda=\lambda\left(  x\text{,
}E^{\prime}\text{, }E\right)  $ in Eq. (\ref{gpsi}) is defined as%
\begin{equation}
\lambda\left(  x\text{, }E^{\prime}\text{, }E\right)  \overset{\text{def}}%
{=}\frac{1}{2}\sqrt{4x^{2}E^{\prime}E+\left(  E^{\prime}-E\right)  ^{2}%
}\text{.}%
\end{equation}
We recall that the states in Eqs. (\ref{originals}) and (\ref{gpsi}) are
expressed in terms of the orthonormal basis $\left\{  \left\vert
w\right\rangle \text{, }\left\vert r\right\rangle \right\}  $ introduced in
Section II.\textbf{ }Substituting Eqs. (\ref{geoline}) and (\ref{gpsi}) into
$d^{2}\left(  t\text{, }\lambda\right)  $ in Eq. (\ref{distanza}), leads to%
\begin{equation}
\frac{d_{\text{MFG}}^{2}\left(  t\text{, }\lambda\right)  }{4}=\left\{
1-\left[  \frac{\left(  1-\lambda\right)  ^{2}\cos^{2}\left(  \frac{\lambda
}{\hslash}t\right)  }{1-2\lambda\left(  1-\lambda\right)  }+\frac{\lambda
^{2}\sin^{2}\left(  \frac{\lambda}{\hslash}t\right)  }{1-2\lambda\left(
1-\lambda\right)  }+\frac{\lambda\left(  1-\lambda\right)  \sin\left(
\frac{2\lambda}{\hslash}t\right)  \cos\left(  \frac{\pi}{2}\frac{E^{\prime}%
+E}{\sqrt{4x^{2}E^{\prime}E+\left(  E^{\prime}-E\right)  ^{2}}}\right)
}{1-2\lambda\left(  1-\lambda\right)  }\right]  \right\}  \text{.}\label{lt2}%
\end{equation}
Finally, proceeding as before, we impose $d^{2}\left(  t\text{, }%
\lambda\right)  $ equal to zero so as to find possible roots $\left\{
\lambda\left(  t\right)  \right\}  $. Then, in order to obtain a well-defined
reversible mapping $\lambda:\left(  0\text{, }t_{\ast}\right)  \ni
t\mapsto\lambda\left(  t\right)  \in\left(  0\text{, }1\right)  $ with\textbf{
}$\lambda(0)=0$ and $\lambda\left(  t_{\ast}\right)  =1$, we observe that it
is necessary to have $t_{\ast}=t_{\text{MFG}}(x)\overset{\text{def}}{=}%
\pi\hslash/(\sqrt{4x^{2}E^{\prime}E+\left(  E^{\prime}-E\right)  ^{2}})$ with
$x\in\left(  0\text{, }1\right)  $ such that%
\begin{equation}
\frac{\pi}{2}\frac{E^{\prime}+E}{\sqrt{4x^{2}E^{\prime}E+\left(  E^{\prime
}-E\right)  ^{2}}}=2n\pi\text{,}\label{nsolvable}%
\end{equation}
with arbitrary $n\in%
\mathbb{N}
$. Unlike Eq. (\ref{solvable}) however, there does not exist any real value
$x\in\left(  0\text{, }1\right)  $ in the modified physical scenario of
interest with $E^{\prime}\gg E$ (that is, $\gamma\gg1$ with $E^{\prime
}\overset{\text{def}}{=}\gamma E$) presented in Section II and originally
proposed by Farhi and Gutmann. Indeed, assuming such a working condition, Eq.
(\ref{nsolvable}) yields $x^{2}\overset{E^{\prime}\gg E}{\approx}\left[
\left(  1-16n^{2}\right)  /64n^{2}\right]  \left(  E^{\prime}/E\right)  $.
Thus, Eq. (\ref{nsolvable}) has no solutions for $x$ belonging to the interval
of interest that yields $d^{2}\left(  t\text{, }\lambda\right)  $ identically
equal to zero. For a discussion on the small energy difference regime, we
refer to Appendix B. In summary, given the impossibility of finding a
well-defined reversible mapping between the geodesic line in Eq.
(\ref{geoline}) and the dynamical trajectory traced by the state vector in Eq.
(\ref{gpsi}), we conclude that the modified Farhi-Gutmann quantum evolution is
not described by a geodesic path on the Bloch sphere.

\begin{figure}[t]
\centering
\includegraphics[width=1\textwidth] {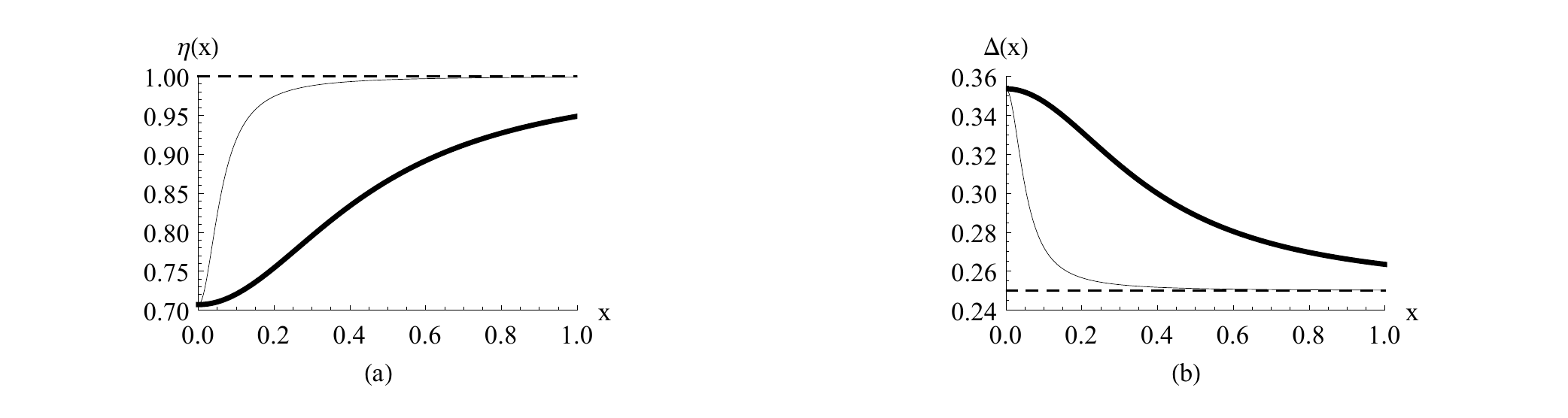}\caption{Part (a): Efficiency $\eta$
versus the quantum overlap $x$ in the case of $\gamma=1$ (dashed line),
$\gamma=1.1$ (thin solid line), and $\gamma=2$ (thick solid line). Part (b):
Uncertainty $\Delta$ versus the quantum overlap $x$ in the case of $\gamma=1$
(dashed line), $\gamma=1.1$ (thin solid line), and $\gamma=2$ (thick solid
line). The original and modified Farhi-Gutmann scenarios are specified by
$\gamma\overset{\text{def}}{=}E^{\prime}/E=1$ and $\gamma>1$, respectively.}%
\label{fig1}%
\end{figure}Finally, by substituting Eq. (\ref{gpsi}) into the efficiency
$\eta$ given in Eq. (\ref{efficiency}) and the time-energy uncertainty
inequality presented in Eq. (\ref{uncertainty}), we obtain%
\begin{equation}
\eta_{\text{MFG}}\left(  x\text{, }E^{\prime}\text{, }E\right)  =\frac
{1}{\sqrt{2}}\left[  \frac{\left(  E^{\prime}-E\right)  ^{2}+4x^{2}E^{\prime
}E}{\left(  E^{\prime}-E\right)  ^{2}+2x^{2}E^{\prime}E}\right]  ^{1/2}<\text{
}1 \label{geff}%
\end{equation}
and,%
\begin{equation}
\Delta_{\text{MFG}}\left(  x\text{, }E^{\prime}\text{, }E\right)
\overset{\text{def}}{=}\left[  \left\langle \Delta E\right\rangle \Delta
t_{\perp}\right]  _{\text{MFG}}=\frac{h}{2\sqrt{2}}\left[  \frac{\left(
E^{\prime}-E\right)  ^{2}+2x^{2}E^{\prime}E}{\left(  E^{\prime}-E\right)
^{2}+4x^{2}E^{\prime}E}\right]  ^{1/2}>h/4\text{,} \label{ua}%
\end{equation}
respectively, for any $E^{\prime}>E$. Thus, by studying the geometry of the
modified Farhi-Gutmann quantum evolution, we arrive at the conclusion that it
does not describe a geodesic motion on the Bloch sphere, is specified by a
non-maximal efficiency $\eta_{\text{MFG}}\left(  x\text{, }E^{\prime}\text{,
}E\right)  $ and the\textbf{ }minimum uncertainty $\Delta_{\text{MFG}}\left(
x\text{, }E^{\prime}\text{, }E\right)  $ is greater than the minimum
achievable value of $h/4$. For the sake of simplicity, we set the Planck
constant $h$ equal to one. Finally, for the sake of clarity, we plot the
efficiency $\eta$ and the uncertainty $\Delta$ as a function of the quantum
overlap $x\in\left(  0\text{, }1\right)  $ for a number of fixed values of the
ratio $\gamma\overset{\text{def}}{=}E^{\prime}/E\geq1$ in Fig. $1$.

\section{Concluding remarks}

In this article, we employed Riemannian geometric concepts to show that time
optimal analog quantum search evolutions between two orthogonal quantum states
are characterized by unit efficiency (see Eq. (\ref{unit})) dynamical
trajectories traced on a projective Hilbert space. In particular, we proved
that these optimal dynamical trajectories are the shortest geodesic paths
joining the initial and the final states of the quantum evolution (see Eqs.
(\ref{lt0}), (\ref{solvable}), and (\ref{lt})). In addition, we verified that
they describe minimum uncertainty evolutions specified by an uncertainty
inequality that is tighter than the ordinary time-energy uncertainty relation
(see Eq. (\ref{unit})). Furthermore, we studied the effects of deviations from
the time optimality condition from our proposed Riemannian geometric
perspective. In particular, by geometric means we found that deviations from
the original Farhi-Gutmann Hamiltonian evolution lead to non-geodesic motion
on the Bloch sphere (see Eqs. (\ref{lt2}) and (\ref{nsolvable})), to
non-maximal efficiency (see Eq. (\ref{geff})) and, finally, to non-minimal
uncertainty of the evolution (see Eq. (\ref{ua})). A summary of our main
results can be visualized in Fig. $1$ and are reported in Table I.

We believe that despite its simplicity, the relevance of our work is
threefold. Firstly, our Riemannian geometric analysis of quantum search
evolutions offers an alternate theoretical perspective on the concept of
optimality with intuitive physical insights arising from familiar concepts
such as shortest path, maximal efficiency, and minimal uncertainty. In this
respect, it becomes especially relevant when taken into consideration together
with Refs. \cite{farhi98,carlophysica19}. Secondly, it could potentially help
providing a practical and systematic way of constructing efficient search
schemes. Indeed, this construction could occur by ranking the maximal
achievable efficiencies of the various search schemes while tuning parameters
of physical relevance that specify the more realistic time-dependent
Hamiltonian at hand \cite{cafaro19,wilczek20}. For instance, it would be of
interest to extend the simple analysis presented here to time-dependent
quantum search Hamiltonians yielding either on-resonance or off-resonance
scenarios \cite{cafaropre20,cafaro20}. In such a case, the set of
experimentally tunable parameters would include, for instance, the energy gap
between two quantum states, the frequency of the external drive field, and the
strength of the external drive field as discussed in Ref. \cite{wilczek20}.
Thirdly, given that realistic quantum algorithms are expected to be both fast
and thermodynamically efficient \cite{castelvecchi97}, our work can be
regarded as a model to emulate in order to find a good geometric measure of
efficiency for thermodynamic processes. Our preliminary results along these
lines have recently appeared in Refs. \cite{cafaropre18,cafaropre20}. Roughly
speaking, the main idea is to replace the geometry of quantum evolutions with
the geometry of thermodynamic processes \cite{ruppeiner95a,quevedo07},
Schrodinger's quantum trajectories with thermal trajectories \cite{brody98}
and, finally, shortest paths on the Bloch sphere with coolest paths on the
manifold of thermal states \cite{diosi96,scandi19,saito20}. Clearly, one may
wonder how to introduce a notion of thermodynamical efficiency in this quantum
searching context. We remark that in the modified scenario with $E^{\prime}\gg
E$, the system moves along a non-geodesic path connecting the initial and
final orthogonal states with a speed $v_{\text{MGF}}=(2/\hslash)\Delta
E_{\text{MFG}}>\pi/\Delta t_{\perp}^{\left(  \text{MFG}\right)  }$. However,
despite exhibiting a speed higher than the one that would specify an evolution
of geodesic type, the modified scenario is not energetically favorable since
the minimal Anandan-Aharonov minimum time-energy uncertainty condition is
violated (that is, $\Delta_{\text{MFG}}>h/4$) with the consequence that the
efficiency as defined in Eq. (\ref{efficiency}) of this particular quantum
mechanical evolution is suboptimal. These considerations, emerging from this
specific physical scenario considered within the framework of quantum search
Hamiltonian evolutions, are reminiscent of the speed-efficiency trade-off
mentioned in our Introduction. At this stage, however, we can only speculate
on the issue of defining a good measure of thermodynamic efficiency in the
context of quantum searching. To be more specific, our next set of explorative
steps in this direction includes the following points: i) a better
understanding of the analogies between quantum mechanical and thermodynamical
relations \cite{brody98,sjoqvist20}; ii) a deeper comprehension of the methods
of thermodynamic geometry employed to identify optimal driving protocols that
minimize the dissipative losses of the underlying thermal processes
\cite{saito20}; iii) an extensive understanding of the geometry of evolutions
of open quantum systems with particular emphasis on the determination of
time-optimal evolutions of impure quantum states \cite{sjoqvist20}; iv) a
quantitative understanding of the possible beneficial effects of dissipation
in quantum searching with the inclusion of thermodynamical arguments
\cite{mizel09}. We believe, for instance, that one of the relevant outcomes of
this cross fertilization between geometry, quantum, and thermal physics will
be a more systematic hybrid method of identifying constructive use of
dissipation in quantum searching. In particular, we expect that the
identification tool will be a measure of efficiency to get to the target state
from a given initial state that is geometrically characterized by a suitable
cost function that ideally maximizes quantum speed and minimizes thermal
dissipation at the same time. Of course, these are mere conjectures at this
point and we shall keep pursuing these fascinating avenues of investigations
in our future scientific efforts.

\begin{table}[t]
\centering
\begin{tabular}
[c]{c|c|c|c}\hline\hline
Quantum Evolution & Motion on Bloch Sphere & Uncertainty of Evolution &
Efficiency of Evolution\\\hline
original Farhi-Gutmann & geodesic & $\Delta=h/4$, minimal & $\eta=1$,
maximal\\
modified Farhi-Gutmann & non-geodesic & $\Delta>h/4$, non-minimal & $\eta<1$,
non-maximal\\\hline
\end{tabular}
\caption{Illustrative representation of the type of motion on the Bloch
sphere, uncertainty of evolution, and efficiency of evolution corresponding to
both the original and the modified Farhi-Gutmann quantum search Hamiltonian
evolutions.}%
\end{table}

\begin{acknowledgments}
C.C. is grateful to the United States Air Force Research Laboratory (AFRL)
Summer Faculty Fellowship Program for providing support for this work. S.R.
acknowledges support from the National Research Council Research Associate
Fellowship program (NRC-RAP). P.M.A. acknowledges support from the Air Force
Office of Scientific Research (AFOSR). Any opinions, findings and conclusions
or recommendations expressed in this material are those of the author(s) and
do not necessarily reflect the views of the Air Force Research Laboratory (AFRL).
\end{acknowledgments}

\pagebreak

\appendix

\section{Geometric efficiency beyond pure states and time-propagators}

In this Appendix, we emphasize several technical details concerning the manner
in which our measure of efficiency $\eta$ in Eq. (\ref{efficiency}) can be
readily extended to more general physical processes.

Our expression for the efficiency Eq.~(\ref{efficiency}) can be written and
interpreted in several forms as shown below%
\begin{equation}
\eta\overset{\text{def}}{=}\frac{s_{0}}{s}=\frac{\int_{\text{geo}%
}ds_{\text{Wootters}}}{\int_{\mathrm{H}}ds_{\text{FS}}}=\frac{2\cos
^{-1}\left[  \left\vert \left\langle \psi\left(  0\right)  |\psi\left(
t_{\ast}\right)  \right\rangle \right\vert \right]  }{2\int_{0}^{t_{\ast}%
}\frac{\Delta E\left(  t^{\prime}\right)  }{\hslash}dt^{\prime}}%
=\frac{2\,\theta_{B}}{2\int_{0}^{t_{\ast}}\frac{\Delta E\left(  t^{\prime
}\right)  }{\hslash}dt^{\prime}}. \label{efficiency:app}%
\end{equation}
The first equality in Eq.~(\ref{efficiency:app}) defines the efficiency $\eta$
as the ratio of two lengths, which by the second equality shows is the ratio
of the Wootters distance along the geodesic connecting the initial and final
states $\left\vert \psi\left(  0\right)  \right\rangle $ and $\left\vert
\psi\left(  t_{\ast}\right)  \right\rangle $ to the the integrated
Fubini-Study distance along the path generated by the Hamiltonian $\mathrm{H}%
$. This later ratio is given by the third equality. As interpreted in the main
text, the last inequality uses the fact that for pure states, the numerator
defines the Bures angle $\theta_{B}$ \cite{carol} via $\cos(\theta_{B}%
)=\sqrt{\mathrm{F}}$ where for pure states $\sqrt{\mathrm{F}}=\left\vert
\left\langle \psi\left(  0\right)  |\psi\left(  t_{\ast}\right)  \right\rangle
\right\vert $ is the Uhlmann Fidelity between the initial and final state
$\left\vert \psi\left(  0\right)  \right\rangle $ and $\left\vert \psi\left(
t_{\ast}\right)  \right\rangle $. The Bures angle is related to the Bures
distance $d_{B}$ \cite{carol} via $d_{B}^{2}=2\,(1-\sqrt{\mathrm{F}}%
)=4\,\sin^{2}(\theta_{B}/2)$, so that we could have also written the numerator
of Eq.~(\ref{efficiency:app}) in terms of this length as $2\,\theta
_{B}=4\,\sin^{-1}\left(  d_{B}/2\right)  $. For infinitesimally close states
this gives $\theta_{B}\approx d_{B}$. Thus, we see that our efficiency
considered as a ratio of geometric lengths is intimately related to the
quantum mechanical concept of the fidelity between the initial and final states.

The concept of fidelity is generalized from pure to mixed states via the
Ulhmann-Jozsa \cite{carol,wilde} fidelity $\mathrm{F}\left[  \rho\left(
0\right)  \text{, }\rho\left(  t_{\ast}\right)  \right]  \overset{\text{def}%
}{=}\left[  \mathrm{tr}\left(  \sqrt{\rho\left(  0\right)  }\rho\left(
t_{\ast}\right)  \sqrt{\rho\left(  0\right)  }\right)  \right]  ^{2}$ which
arises from the overlap of the pure state $\left\vert \psi_{\rho}\left(
0\right)  \right\rangle $ and $\left\vert \psi_{\rho}\left(  t_{\ast}\right)
\right\rangle $ \emph{purifications} of the initial and final states $\rho(0)$
and $\rho(t_{\ast})$, maximized over an arbitrary unitary in the higher
dimensional purification Hilbert space. (Note that: $\mathrm{tr}%
_{R}[\left\vert \psi_{\rho}\right\rangle \left\langle \psi_{\rho}\right\vert
]=\rho$, where the purified state $\left\vert \psi_{\rho}\right\rangle $ lives
in the composite Hilbert space $\mathcal{H}_{R}\otimes\mathcal{H}_{S}$ of
system-$S$ ($\rho$) and reservoir-$R$). The Bures angle and Bures distance
retain their pure-state form \cite{carol}, i.e. $\cos(\theta_{B}%
)=\sqrt{\mathrm{F}}\left[  \rho\left(  0\right)  \text{, }\rho\left(  t_{\ast
}\right)  \right]  $ and $d_{B}^{2}=2\,\left(  1-\sqrt{\mathrm{F}}\left[
\rho\left(  0\right)  \text{, }\rho\left(  t_{\ast}\right)  \right]  \right)
$. Note that the fidelity is a \emph{total} distance in the sense that its
computation only relies upon the knowledge of the state at either end of the
geodesic that connects the initial and final state. One might ask if there is
some differential quantity for which the fidelity is the integrated version
along the geodesic. The answer is yes, and this quantity is the Quantum Fisher
Information (QFI). This leads to a new interpretation of the denominator in
Eq.~(\ref{efficiency:app}).

Let us note that $\mathcal{F}_{Q}\left(  t\right)  \overset{\text{def}}%
{=}\mathrm{tr}\left[  \rho\left(  t\right)  L^{2}\left(  t\right)  \right]  $
denotes the quantum Fisher information for time estimation along the
trajectory specified by the system evolution, and $L\left(  t\right)  $ is the
so-called symmetric logarithmic derivative operator defined in an implicit
fashion by the equation $d\rho/dt=\left[  \rho\left(  t\right)  L\left(
t\right)  +L\left(  t\right)  \rho\left(  t\right)  \right]  /2$
\cite{brau94}. Moreover, the connection between $\mathcal{F}_{Q}\left(
t\right)  $ in the denominator of $\eta$ in Eq. (\ref{efficiency:app}) and the
dispersion $\Delta E\left(  t\right)  $ of the Hamiltonian operator \textrm{H}
in the denominator of $\eta$ in Eq. (\ref{efficiency}) can be made transparent
by observing that the analogue of $\left\vert \left\langle \psi\left(
t\right)  |\psi\left(  t+dt\right)  \right\rangle \right\vert ^{2}=1-\left[
\Delta E^{2}\left(  t\right)  /\hslash^{2}\right]  dt^{2}+O\left(
dt^{3}\right)  $ is $\mathrm{F}\left(  t\text{, }t+dt\right)  =1-\left[
\mathcal{F}_{Q}\left(  t\right)  /4\right]  dt^{2}+O\left(  dt^{3}\right)  $
for mixed quantum states \cite{taddei13}. Therefore, the square root
$\sqrt{\mathcal{F}_{Q}\left(  t\right)  }$ of the quantum Fisher information
replaces $2\left[  \Delta E\left(  t\right)  /\hslash\right]  $ for pure
states and is generally proportional to the instantaneous speed of separation
between two infinitesimally closed mixed quantum states. This allows us to
write the efficiency in terms of the fidelities as%
\begin{equation}
\eta\overset{\text{def}}{=}\frac{s_{0}}{s}=\frac{2\cos^{-1}\left(
\sqrt{\mathrm{F}}[\rho(0),\rho(t_{\ast})]\right)  }{\int_{0}^{t^{\ast}}%
\sqrt{\mathcal{F}_{Q}(t^{\prime})}\,dt^{\prime}}, \label{efficiency:app:2}%
\end{equation}
which now holds in general for mixed states.

Secondly, the Hamiltonian operator \textrm{H }and the temporal parameter $t$
can be replaced by any other Hermitian operator \textrm{A}$_{\xi}$ (for
instance, the number operator, the momentum operator, or the angular momentum
along the quantization axis) and any arbitrary parameter $\xi$ (for instance,
the phase of a clock or the strength of an external field), respectively. The
parameter $\xi$ describes the evolution of the physical system by the action
of the unitary operator $U_{\mathrm{A}_{\xi}}\left(  \xi\right)
\overset{\text{def}}{=}e^{\frac{i}{\hslash}\xi\mathrm{A}_{\xi}}$ which
replaces the usual Schr\"{o}dinger time-propagator. As a consequence, the
usual Anandan-Aharonov time-energy uncertainty inequality, $\left\langle
\Delta E\right\rangle \Delta t_{\perp}\geq h/4$, can be generalized to assume
the form $\Delta\mathrm{A}_{\xi}\delta\xi\geq h/4$ with $\delta\xi$ being
essentially the precision with which $\xi$ can be determined \cite{mann98}. In
addition, we remark that for pure states the quantum Fisher information is a
multiple of the variance of $\mathrm{A}_{\xi}$. For mixed states, instead, the
variance provides only an upper bound on the Fisher information \cite{boixo07}%
. Therefore, given this intimate connection between the quantum Fisher
information and the variance of the Hermitian generator $\mathrm{A}_{\xi}$ of
the displacements in $\xi$, the usual Anandan-Aharonov time-energy uncertainty
inequality can be regarded as being replaced by a generalized uncertainty
relation, $\delta\xi\geq\left(  h/2\right)  \mathcal{F}_{\xi}^{-1/2}\left(
t\right)  $, that derives from the Cramer-Rao bound that appears in precision
quantum metrology \cite{milburn96,giovannetti06,gerry20}.

Lastly, it should be noted that along the geodesic, i.e. the shortest distance
connecting the initial and final states, we have $\int_{0,\,\text{geo}%
}^{t^{\ast}}\sqrt{\mathcal{F}_{Q}(t^{\prime})}\,dt^{\prime}=\sqrt{\mathrm{F}%
}[\rho(0),\rho(t_{\ast})]$ so that the QFI is the infinitesimal version of the
quantum fidelity. Further, along any longer (non-geodesic) path ($s>s_{0}$)
generated by a Hamiltonian $\mathrm{H}$, we have $\int_{0,\,H}^{t^{\ast}}%
\sqrt{\mathcal{F}_{Q}(t^{\prime})}\,dt^{\prime}\leq\sqrt{\mathrm{F}}%
[\rho(0),\rho(t_{\ast})]$. This allows us to generalize the concept of
efficiency $0\leq\eta\rightarrow\tilde{\eta}\leq1$ to a quantity involving the
only ratio of fidelities and/or of integrated QFIs along the optimal geodesic
($s_{0}$) and the evolved (under $\mathrm{H}$) path ($s>s_{0}$)
\begin{equation}
\eta=\frac{s_{0}}{s}\longleftrightarrow\tilde{\eta}\overset{\text{def}}%
{=}\frac{\int_{0,\,\mathrm{H}}^{t^{\ast}}\sqrt{\mathcal{F}_{Q}(t^{\prime}%
)}\,dt^{\prime}}{\sqrt{\mathrm{F}}[\rho(0),\rho(t_{\ast})]_{\text{geo}}}%
=\frac{\int_{0,\,\mathrm{H}}^{t^{\ast}}\sqrt{\mathcal{F}_{Q}(t^{\prime}%
)}\,dt^{\prime}}{\int_{0,\,\text{geo}}^{t^{\ast}}\sqrt{\mathcal{F}%
_{Q}(t^{\prime})}\,dt^{\prime}}. \label{efficiency:tilde:app}%
\end{equation}
Both measures of efficiencies $\eta$ and $\tilde{\eta}$ quantify the same
concepts, in terms of inverse ratios, relating the initial and final states of
the system: (i) the geometric point of view: $\eta=s_{0}/s\leq1$, i.e. the
length along the path generated by $\mathrm{H}$ is \emph{greater} than the
optimal (shortest) geodesic path, and (ii) the fidelity/QFI point of view:
$\tilde{\eta}\leq1$, i.e. the fidelity, or integrated QFI, along the path
generated by $\mathrm{H}$ is \emph{less} than that of along the optimal path.

The study of various geometric characteristics along evolution of density
operators is becoming increasingly important and deserves special care
\cite{sjoqvist20}. For this reason, we leave the quantitative analysis of the
physical usefulness of the efficiency measures $\eta$ and $\tilde{\eta}$ in
Eq. (\ref{efficiency:app}) (and Eq. (\ref{efficiency:app:2})) and Eq.
(\ref{efficiency:tilde:app}) in analog quantum searching and precision
metrology to forthcoming efforts.

\section{Small energy difference regime}

In this Appendix, we comment for the sake of mathematical completeness on the
geodesic constraint equation $d^{2}\left(  t\text{, }\lambda\right)  =0$ in
the case of the small energy difference regime, although our main focus in the
manuscript is devoted to the large energy difference regime specified by
$E^{\prime}\gg E$.

When we relax the working condition $E^{\prime}\gg E$ and consider the low
energy difference scenario where $E^{\prime}\overset{\text{def}}{=}\gamma E$
and $E$ are sufficiently close with $E^{\prime}>E$, imposing $d^{2}\left(
t\text{, }\lambda\right)  $ in Eq. (\ref{lt2}) to be equal to zero requires
that the quantum overlap $x$ satisfies the condition $x^{2}=x^{2}\left(
n\text{, }\gamma\right)  $ where,%
\begin{equation}
x^{2}\left(  n\text{, }\gamma\right)  \overset{\text{def}}{=}\frac{1}{64\gamma
n^{2}}\left[  \left(  1-16n^{2}\right)  \gamma^{2}+\left(  2+32n^{2}\right)
\gamma+\left(  1-16n^{2}\right)  \right]  \text{,}\label{heart}%
\end{equation}
with $n\in%
\mathbb{N}
$ and $\gamma>1$. A simple calculation shows that $x^{2}$ in\ Eq.
(\ref{heart}) assumes positive values on the set $\mathcal{I}_{n}%
\overset{\text{def}}{=}\left[  i_{-}\left(  n\right)  \text{, }i_{+}\left(
n\right)  \right]  $ with $i_{\pm}\left(  n\right)  \overset{\text{def}}%
{=}\left(  32n^{2}\pm16n+2\right)  /(32n^{2}-2)$.\textbf{ }However,
$\mathcal{I}_{n}$ is a set whose measure vanishes asymptotically since\textbf{
}$\mu\left(  \mathcal{I}_{n}\right)  \overset{\text{def}}{=}16n/(16n^{2}%
-1)\overset{n\gg1}{\approx}1/n\rightarrow0$ when\textbf{ }$n$\textbf{
}approaches infinity. In summary, the set\textbf{\ }$\mathcal{I}_{n}%
$\textbf{\ }tend to shrink and eventually, vanish. Moreover, the set
$\mathcal{I}_{n}$ contains elements that violate the condition $\gamma>1$.
Indeed, $i_{-}\left(  n\right)  >1$ if and only if $n<1/4$. Clearly, this is
impossible since $n\in%
\mathbb{N}
$. In particular, for any $\gamma>i_{+}\left(  1\right)  \overset{\text{def}%
}{=}5/3$ with $5/3$ being the upper bound of the set $\mathcal{I}_{n}$ with
the largest measure, that is $\mathcal{I}_{1}$ with $\mu\left(  \mathcal{I}%
_{1}\right)  =16/15$, $x^{2}$ in Eq. (\ref{heart}) becomes negative. Thus, we
can conclude that Eq. (36) has no solution $x$ that belongs to the
interval\ $\left(  0\text{, }1\right)  $\ for any real\ $\gamma>i_{+}\left(
1\right)  $. Interestingly, the limit of large $n$ values can also be
physically motivated. Indeed, from a physics standpoint, we expect
$t_{\text{MFG}}$ to become very large when $E^{\prime}\gtrsim E$ in the study
of the quantum mechanical evolution between nearly orthogonal quantum states
with nearly zero quantum overlap since $t_{\text{MFG}}$ is inversely
proportional to the energy level separation of the system and this equals
$\left[  4x^{2}\gamma+\left(  1-\gamma\right)  ^{2}\right]  E$
\cite{cafaro19,gassner20}. Imposing the condition expressed in Eq.
(\ref{nsolvable}), the long time limit is recovered when $n\gg1$ since such a
condition requires $t_{\text{MFG}}$ to be proportional to $n$ with constant of
proportionality coefficient given by\textbf{ }$2h/(E^{\prime}+E)$\textbf{. }As
a side remark, this long time limit is reminiscent of the infinite temporal
duration of highly efficient ideal reversible thermodynamic processes that
occur in the absence of dissipation. In summary, we can safely conclude from
our discussion that $x\notin\left(  0\text{, }1\right)  $\ for any positive
integer $\gamma\in%
\mathbb{Z}
_{+}$\ with $\gamma>1$\ in any of the two energetic regimes (that is,
$E^{\prime}\gg E$ and $E^{\prime}>E$) of the modified quantum search scenario
if $d^{2}\left(  t\text{, }\lambda\right)  $ in Eq. (\ref{lt2}) is required to
be identically zero.
\end{document}